\documentclass[12pt]{article}
\usepackage{graphicx}
\input xy
\xyoption{all}

\usepackage{graphicx}

\begin{document}

\hfill VPI-IPNAS-08-02

% Will need to email Tatsu to get a preprint number. 

\vspace{1.5in}

\begin{center}
{\large\bf Recent developments in heterotic compactifications}

\vspace{0.1in}

Eric Sharpe \\

$\,$

Physics Department\\
Robeson Hall (0435) \\
Virginia Tech\\
Blacksburg, VA  24061\\

{\tt ersharpe@vt.edu}

$\,$

\end{center}

In this short review,
we outline three sets of developments in understanding heterotic string
compactifications.  First, we outline recent progress in heterotic analogues
of quantum cohomology computations.  Second, we discuss a potential swampland
issue in heterotic strings, and new heterotic string constructions that can
be used to fill in the naively missing theories.  Third, we discuss
recent developments in string compactifications on stacks and their
applications, concluding with an outline of work-in-progress on heterotic
string compactifications on gerbes.

Contribution to the proceedings of the Virginia Tech Sowers workshop, May 2007.

\begin{flushleft}
January 2008
\end{flushleft}

\newpage

\tableofcontents

%\newpage

\section{Introduction}

Over the last several years, heterotic strings have been undergoing something
of a revival.  There has been a lot of interest in heterotic strings
on non-K\"ahler manifolds (see {\it e.g.} 
\cite{b2a,b2b,b2c,b2d,allan1,allan2}), on MSSM derivations from string
theory \cite{penn1,penn2}, as well as many other matters related
to heterotic strings, 
as reflected in Volker Braun's, Savdeep Sethi's, and Li-Sheng Tseng's talks
at this workshop\footnote{Presented at the Virginia Tech Sowers workshop,
May 2007.}.

In this talk,
we shall outline three other recent developments in heterotic
strings.

First, we shall discuss recent progress on understanding nonperturbative
corrections in heterotic strings, the heterotic analogue of curve
corrections and quantum cohomology.  These will be computed by a heterotic
analogue of the two-dimensional topological field theory known as the A model.
We will briefly review the A model at the same time as we present its
heterotic generalization.  This work was motivated by efforts to
understand the heterotic generalization of mirror symmetry.

Second, we shall discuss a potential heterotic swampland, referring
to the fact that many $E_8$ bundles with connection cannot be described
by the standard heterotic worldsheet construction.  We will outline
new heterotic worldsheet CFT constructions which will make it possible
to describe all the $E_8$ bundles with connection.  Although there will
not be a heterotic swampland of this form, this does serve as a warning on
the dangers of performing statistical computations within fixed worldsheet
constructions.

Finally, we shall discuss the recent understanding of string compactifications
on stacks.  We shall also outline descriptions of some of those
compactifications with gauged linear sigma models.  Understanding
string compactifications on stacks not only makes predictions for
{\it e.g.} certain Gromov-Witten invariants, but it also yields insight into
ordinary-seeming gauged linear sigma models.  We shall briefly outline the
analysis of the GLSM for the complete intersection Calabi-Yau
${\bf P}^7[2,2,2,2]$, which has at its Landau-Ginzburg point another
Calabi-Yau given by a branched double cover of ${\bf P}^3$.  The branched
double cover appears physically in a novel way, and furthermore is not 
birational to the original complete intersection ${\bf P}^7[2,2,2,2]$,
contradicting some of the lore on gauged linear sigma models.
Finally, we shall outline how heterotic string compactifications on
special stacks known as gerbes appear to provide new examples of CFT's.

\section{Nonperturbative corrections in heterotic strings}

In this section, we shall outline the results of
\cite{sharpe02a,sharpe02c} on nonperturbative corrections to heterotic
string compactifications.  See \cite{sharpe02b} for another
review, and \cite{ade,kg,ilarion1,ilarion2,gs1,gs2} for more recent results.

Roughly, there are two sources of nonperturbative corrections in heterotic
strings:
\begin{itemize}
\item gauge instantons and five-branes, and
\item worldsheet instantons -- from strings wrapping minimal area 2-cycles
(``holomorphic curves'') in spacetime.
\end{itemize}
In this talk, we shall focus on the latter class, in perturbative
worldsheet theories.

Worldsheet instantons generate superpotential terms in the target-space
effective field theory.  For example, for a heterotic theory with a rank
three bundle, breaking an $E_8$ to $E_6$, there are
\begin{itemize}
\item ${\bf \overline{27} }^3$ couplings -- on the (2,2) locus,
{\it i.e.} when the gauge bundle is the same as the tangent bundle,
these are computed by A model correlation functions.
\item ${\bf 27}^3$ couplings -- on the (2,2) locus, these are computed by
B model correlation functions.
\item Singlet couplings, such as potential terms lifting moduli -- these
are not computed on the (2,2) locus by any topological field theory,
and so in principle are harder to compute directly.
See \cite{evaed,candelasetal,bw} for
some (vanishing) results concerning such singlet couplings.
\end{itemize}

In this talk, we will focus on the first two classes of superpotential
terms.  Off the (2,2) locus, {\it i.e.} when the gauge bundle is not 
the same as the tangent bundle,
there exist analogues of the A and B models.
These are no longer strictly topological field theories, though they
become topological field theories on the (2,2) locus.
Nevertheless, although they are not quite the same as topological field
theories, they have many of the same properties as topological field
theories, and in particular some correlation functions still have a 
mathematical understanding.

These quasi-topological field theories also have some unusual symmetries;
in particular,
the (0,2) A model on a space $X$ with gauge bundle ${\cal E}$
is isomorphic to the (0,2) B model on $X$ with gauge bundle ${\cal E}^{\vee}$.
(For example, this means the (2,2) A model on $X$ is the same as the
(0,2) B model on $X$ with gauge bundle $T^*X$ instead of $TX$.) 

In addition to computing superpotential terms, these quasi-topological
field theories also have applications to understanding the
(0,2) generalization of mirror symmetry.
Recall that ordinary mirror symmetry exchanges pairs of topologically-distinct
spaces, dualizing quantum-corrected computations into classical computations,
and exchanging cohomology, in the sense that if $X$ and $Y$ are mirror,
then $h^{p,q}(X) = h^{d-p,q}(Y)$ where $d$ is the dimension of $X$.

In principle, (0,2) mirror symmetry exchanges spaces together with bundles.
Instead of swapping ordinary cohomology classes, it exchanges sheaf
cohomology groups:  if $(X_1, {\cal E}_1)$ is exchanged with
$(X_2, {\cal E}_2)$, then $H^1(X_1,{\cal E}_1)$ is exchanged with
$H^1(X_2, {\cal E}_2^{\vee})$.  In the special case that ${\cal E}_i = TX_i$,
(0,2) mirror symmetry reduces to ordinary mirror symmetry.

At the present time, (0,2) mirror symmetry is not well-understood.
One example of evidence for (0,2) mirror symmetry is numerical computations
\cite{rsw}, in which sheaf cohomology classes of a large
number of examples were computed.  When one graphs the set of dimensions of
sheaf cohomology groups, the graph is symmetric.
Other work on the subject can be found in \cite{rsw,rs,r2,r3,abs}.

In particular, \cite{abs} proposed that there should exist a (0,2)
analogue of `quantum cohomology' computations, encoding worldsheet
instanton corrections, as would be computed ordinarily in the A model
topological field theory.  The papers \cite{sharpe02a,sharpe02c,sharpe02b}
and others since ({\it e.g.} \cite{ade,kg,ilarion1,ilarion2,gs2})
have begun building the details of those proposed
(0,2) quantum cohomology computations, and that is what we shall discuss
in this section.

Since ordinary quantum cohomology rings are operator product rings in the
A model topological field theory, we shall discuss the (0,2) analogue
of the A model.  First, 
let us first review the ordinary A model.
This is a two-dimensional quantum field theory with lagrangian
\begin{displaymath}
g_{i \overline{\jmath}} \overline{\partial} \phi^i \partial 
\phi^{\overline{\jmath}} \: + \:
i g_{i \overline{\jmath}} \psi_-^{\overline{\jmath}} D_z \psi_-^i \: + \:
i g_{i \overline{\jmath}} \psi_+^{\overline{\jmath}} 
D_{\overline{z}} \psi_+^i \: + \:
R_{i \overline{\jmath} k \overline{l}} \psi_+^i \psi_+^{\overline{\jmath}} 
\psi_-^k
\psi_-^{\overline{l}}
\end{displaymath}
where the $\phi$ are maps from the worldsheet $\Sigma$ into a target space $X$,
and the Grassmann field $\psi$ couple to bundles as follows:
\begin{displaymath}
\begin{array}{cc}
\psi_-^i ( \equiv \chi^i ) \: \in \: \Gamma( ( \phi^* T^{0,1} X)^{\vee}) &
\psi_+^i ( \equiv \psi_z^i) \: \in \: \Gamma( K \otimes \phi^* T^{1,0} X) \\
\psi_-^{\overline{\imath}} ( \equiv \psi_{\overline{z}}^{\overline{\imath}} ) \:
 \in \:
\Gamma( \overline{K} \otimes \phi^* T^{0,1} X ) &
\psi_+^{\overline{\imath}} ( \equiv \chi^{\overline{\imath}} ) \: \in \:
\Gamma( (\phi^* T^{1,0} X)^{\vee}).
\end{array}
\end{displaymath}
These fields are not worldsheet fermions, but rather are worldsheet vectors
and scalars -- a result of the topological twisting.
Half of the original supersymmetries between worldsheet scalars, forming
the scalar supercharge or BRST operator, 
and under the action of that supercharge,
\begin{displaymath}
\begin{array}{cc} 
\delta \phi^i \:  \propto \: \chi^i, &
\delta \phi^{\overline{\imath}} \: \propto \: \chi^{\overline{\imath}} \\
\delta \chi^i \: = \: 0, & \delta \chi^{\overline{\imath}} \: = \: 0 \\
\delta \psi_z^i \: \neq \: 0, &  \delta \psi_{\overline{z}}^{\overline{\imath}}
\: \neq \: 0.
\end{array}
\end{displaymath}
As a result, the 
BRST-invariant worldsheet scalar states are built from products of
$\chi$'s, and there is a well-known isomorphism to cohomology given by
\begin{displaymath}
\begin{array}{rcl}
{\cal O} \: \sim \: b_{i_1 \cdots i_p \overline{\imath}_1 \cdots \overline{\imath}_q }
\chi^{\overline{\imath}_1} \cdots \chi^{\overline{\imath}_q} \chi^{i_1} \cdots \chi^{i_p}
& \leftrightarrow & H^{p,q}(X), \\
Q & \leftrightarrow & d.
\end{array}
\end{displaymath}

Next, let us examine comparable facts about the (0,2) analogue of the
A model.  This is based on the heterotic lagrangian
\begin{displaymath}
g_{i \overline{\jmath}} \overline{\partial} \phi^i \partial \phi^{\overline{\jmath}} \: + \:
i h_{a \overline{b}} \lambda_-^{\overline{b}} D_z \lambda_-^a \: + \:
i g_{i \overline{\jmath}} \psi_+^{\overline{\jmath}} D_{\overline{z}} \psi_+^i \: + \:
F_{i \overline{\jmath} a \overline{b}} \psi_+^i \psi_+^{\overline{\jmath}} \lambda_-^a
\lambda_-^{\overline{b}}
\end{displaymath}
describing a heterotic string propagating on a space $X$ with
left-movers coupling to a holomorphic vector bundle ${\cal E}$,
in which the Grassmann fields $\psi$, $\lambda$ couple to bundles as follows:
\begin{displaymath}
\begin{array}{cc}
\lambda_-^a  \: \in \: \Gamma( \phi^* \overline{\cal E}) &
\psi_+^i  \: \in \: \Gamma( K \otimes \phi^* T^{1,0} X) \\
\lambda_-^{\overline{b}}  \: \in \:
\Gamma( \overline{K} \otimes \phi^* \overline{\cal E} ) &
\psi_+^{\overline{\imath}}  \: \in \:
\Gamma( (\phi^* T^{1,0} X)^{\vee}).
\end{array}
\end{displaymath}
As before, the $\psi$ and $\lambda$ are no longer worldsheet fermions,
but rather are scalars and vectors.  Because of the asymmetry between left-
and right-moving fields, this theory must satisfy the conditions
\begin{displaymath}
\Lambda^{top} {\cal E}^{\vee} \cong K_X, \: \: \:
\mbox{ch}_2({\cal E}) = \mbox{ch}_2(TX).
\end{displaymath}
The second of these is the well-known anomaly cancellation condition of
perturbative heterotic strings, the first is another condition present
only in the twisted theory -- a close analogue of the anomaly in the 
closed string B model that makes it well-defined only for complex
K\"ahler manifolds obeying $K_X^{\otimes 2}$ trivial \cite{sharpe02c}.

Here, the BRST-invariant worldsheet scalar states that one considers 
are products of $\psi$'s, $\lambda$'s, and are in one-to-one correspondence
with elements of sheaf cohomology groups
\begin{displaymath}
{\cal O} \: \sim \:
b_{\overline{\imath}_1 \cdots \overline{\imath}_n a_1 \cdots a_p}
\psi_+^{\overline{\imath}_1} \cdots \psi_+^{\overline{\imath}_n}
\lambda_-^{a_1} \cdots \lambda_-^{a_p}
\: \leftrightarrow \:
H^n(X, \Lambda^p {\cal E}^{\vee} ).
\end{displaymath}

In the special case that ${\cal E} = TX$, this pseudo-topological field
theory becomes the A model, a true topological field theory.
(For example, 
\begin{displaymath}
H^q(X, \Lambda^p(TX)^{\vee})
= H^{p,q}(X)
\end{displaymath}
so the state counting specializes in the desired fashion.)

Next, let us turn to correlation function computations in these theories.

In the A model, the classical contributions to correlation functions
are computed as follows.
For $X$ compact and of dimension $n$, there are $n$
$\chi^i$, $\chi^{\overline{\imath}}$ zero modes, plus bosonic zero modes
whose moduli space is $X$ itself, so the correlation function reduces to
\begin{displaymath}
< {\cal O}_1 \cdots {\cal O}_m > \: = \:
\int_X H^{p_1, q_1}(X) \wedge \cdots \wedge H^{p_m, q_m}(X).
\end{displaymath}
(In our notation, we schematically indicate representatives of cohomology
classes by writing the cohomology groups themselves.)
In addition, there is a selection rule from the left and right $U(1)_R$
symmetries that says the classical contributions to correlation functions are 
only nonzero when
\begin{displaymath}
\sum_i p_i = \sum_i q_i = n.
\end{displaymath}
Putting this together, we see that
the classical contribution has the form
\begin{displaymath}
< {\cal O}_1 \cdots {\cal O}_m > \: \sim \:
\int_X (\mbox{top-form} ).
\end{displaymath}

Next, let us consider the analogous computation in the (0,2) analogue of
the A model.
For $X$ compact of dimension $n$, and ${\cal E}$ of rank $r$,
there are $n$ $\psi_+^{\overline{\imath}}$ zero modes and $r$ $\lambda_-^a$
zero modes, so the classical contribution to a correlation function
is of the form
\begin{displaymath}
< {\cal O}_1 \cdots {\cal O}_m > \: = \:
\int_X H^{q_1}(X, \Lambda^{p_1} {\cal E}^{\vee} ) \wedge \cdots \wedge
H^{q_m}(X, \Lambda^{p_m} {\cal E}^{\vee} ).
\end{displaymath}

As before, there is a selection rule, which now enforces
\begin{displaymath}
\sum_i q_i = n, \: \: \:
\sum_i p_i = r
\end{displaymath}
for classical contributions.
Therefore, the classical contributions have the form
\begin{displaymath}
< {\cal O}_1 \cdots {\cal O}_m > \: \sim \:
\int_X H^{top} (X, \Lambda^{top} {\cal E}^{\vee} ).
\end{displaymath}
The constraint $\Lambda^{top} {\cal E}^{\vee} \cong K_X$ makes the
integrand a top-form.

Next, let us turn to a non-classical contribution to a correlation function.
Again let us first consider the original A model before studying the (0,2)
analogue.
In the standard A model, the moduli space of bosonic zero modes
in some non-classical sector will be a moduli space ${\cal M}$ of
worldsheet instantons.
If there are no $\psi_z^i$ or $\psi_{\overline{z}}^{\overline{\imath}}$
zero modes,
then the contribution to a correlation function is of the form
\begin{displaymath}
< {\cal O}_1 \cdots {\cal O}_m > 
\: \sim \: 
\int_{ {\cal M} }
H^{p_1,q_1}({\cal M}) \wedge \cdots
\wedge H^{p_m,q_m}({\cal M}).
\end{displaymath}
More generally, taking into account the possibility of
$\psi_z^i$ or $\psi_{\overline{z}}^{\overline{\imath}}$
zero modes,
the contribution to a correlation function will be of the form
\begin{displaymath}
< {\cal O}_1 \cdots {\cal O}_m > 
\: \sim \: 
\int_{ {\cal M} }
H^{p_1,q_1}({\cal M}) \wedge \cdots
\wedge H^{p_m,q_m}({\cal M})
\wedge c_{top}(\mbox{Obs}).
\end{displaymath}
In all cases, the contribution has the form
\begin{displaymath}
< {\cal O}_1 \cdots {\cal O}_m > 
\: \sim \: 
\int_{ {\cal M} }
(\mbox{top form}).
\end{displaymath}

Next, let us consider the analogous computations to the (0,2) analogue of
the A model.
Here, the bundle ${\cal E}$ on $X$ induces a holomorphic vector bundle\footnote{
In general, this will only be a sheaf.  However, for the bundle constructions
under consideration, over toric varieties, the induced sheaf will always
be locally free.}
${\cal F}$ of $\lambda$ zero modes on
the moduli space ${\cal M}$.  Mathematically,
\begin{displaymath}
{\cal F} \equiv R^0 \pi_* \alpha^* {\cal E}
\end{displaymath}
where
\begin{displaymath}
\pi: \Sigma \times {\cal M} \rightarrow {\cal M}, \: \: \:
\alpha: \Sigma \times {\cal M} \rightarrow X.
\end{displaymath}
On the (2,2) locus, where ${\cal E} = TX$, one has ${\cal F} = T {\cal M}$.
When there are no `excess' (worldsheet vector) zero modes, the contribution
to the correlation function is of the form
\begin{displaymath}
< {\cal O}_1 \cdots {\cal O}_m > \: \sim \: \int_{ {\cal M} } H^{top}(
{\cal M}, \Lambda^{top} {\cal F}^{\vee} ).
\end{displaymath}
When we apply the physical consistency conditions and the
Grothendieck-Riemann-Roch theorem, we find
\begin{displaymath}
\left. \begin{array}{c}
\Lambda^{top} {\cal E}^{\vee} \cong K_X \\
\mbox{ch}_2({\cal E}) = \mbox{ch}_2(TX)
\end{array} \right\} \: \stackrel{GRR}{\Longrightarrow} \:
\Lambda^{top} {\cal F}^{\vee} \cong K_{ {\cal M} }
\end{displaymath}
so again the integrand is a top-form.
In the general case, one has
\begin{displaymath}
\begin{array}{rcl}
< {\cal O}_1 \cdots {\cal O}_m > & \sim &
\int_{ {\cal M} } H^{\sum q_i}\left(
{\cal M}, \Lambda^{\sum p_i} {\cal F}^{\vee} \right) \wedge \\
& & \hspace*{0.5in}
H^n\left( {\cal M}, \Lambda^n {\cal F}^{\vee} \otimes \Lambda^n {\cal F}_1 
\otimes \Lambda^n (\mbox{Obs})^{\vee} \right)
\end{array}
\end{displaymath}
where
\begin{displaymath}
\begin{array}{cc}
\psi_+^{\overline{\jmath}} \sim T {\cal M} = R^0 \pi_* \alpha^* TX &
\lambda_-^a \sim {\cal F} = R^0 \pi_* \alpha^* {\cal E} \\
\psi_+^i \sim \mbox{Obs} = R^1 \pi_* \alpha^* TX &
\lambda_-^{\overline{b}} \sim {\cal F}_1 \equiv R^1 \pi_* \alpha^* {\cal E}.
\end{array}
\end{displaymath}
Applying anomaly constraints as before, we find
\begin{displaymath}
\left.
\begin{array}{c}
\Lambda^{top} {\cal E}^{\vee} \cong K_X \\
\mbox{ch}_2 ({\cal E}) = \mbox{ch}_2(TX) 
\end{array} \right\}
 \: \stackrel{GRR}{\Longrightarrow} \:
\Lambda^{top} {\cal F}^{\vee} \otimes \Lambda^{top} {\cal F}_1
\otimes \Lambda^{top} ( \mbox{Obs} )^{\vee} \: \cong \: K_{ {\cal M} }
\end{displaymath}
so, again, the integrand is a top-form.
On the (2,2) locus, this reduces to the previous result via Atiyah classes
\cite{atiyah}. 

To do any computations with these general formulas, we need explicit expressions
for the space ${\cal M}$ and bundle ${\cal F}$.
Luckily, the gauged linear sigma model naturally provides such expressions.
Gauged linear sigma models are two-dimensional gauge theories,
generalizations of the ${\bf C} {\bf P}^n$ model, which are important
in string compactifications.  They renormalization-group flow to 
nonlinear sigma models and other conformal field theories, and
technical questions about the CFT's will (often) become easier questions
about the gauged linear sigma model.  For example, the worldsheet instantons
of the IR nonlinear sigma model are the two-dimensional gauge instantons of the
gauge theory.

To be specific, let us consider the example of ${\bf C} {\bf P}^{N-1}$.
Physically, this is described by $N$ chiral superfields $x_1, \cdots, x_N$,
each of charge 1.
For degree $d$ maps, and a genus zero worldsheet,
we expand in a basis of zero modes:
\begin{displaymath}
x_i \: = \: x_{i0} u^d \: + \:
x_{i1}u^{d-1}v \: + \: \cdots \: + \:
x_{id} v^d
\end{displaymath}
where $u, v$ are homogeneous coordinates on the worldsheet ${\bf P}^1$.
Taking the $(x_{ij})$ to be homogeneous coordinates
on ${\cal M}$ \cite{daveronen}, we find
${\cal M} = {\bf P}^{N(d+1)=1}$.

We can do something very similar to build ${\cal F}$.
For example, suppose ${\cal E}$ is a completely reducible bundle:
\begin{displaymath}
{\cal E} \: = \: \oplus_a {\cal O}(
\vec{n}_a).
\end{displaymath}
Expanding the left-moving fermions in a basis of zero modes,
on a genus zero worldsheet, we find:
\begin{displaymath}
\lambda_-^a \: = \:
\lambda_-^{a0} u^{\vec{n}_a \cdot \vec{d} + 1} \: + \:
\lambda_-^{a1} u^{\vec{n}_a \cdot d} v
\: + \: \cdots.
\end{displaymath}
We then identify each $\lambda_-^{ai}$ with 
${\cal O}(\vec{n}_a)$ on ${\cal M}$.
Thus, in this case,
\begin{displaymath}
{\cal F} \: = \:
\oplus_a H^0\left({\bf P}^1,
{\cal O}(\vec{n}_a \cdot \vec{d}) \right)
\otimes_{ {\bf C} } {\cal O}(\vec{n}_a).
\end{displaymath}

There are analogous expressions for more general constructions appearing
in (0,2) gauged linear sigma models, see \cite{sharpe02a}.

Next, let us quickly review ordinary quantum cohomology, before describing
the (0,2) analogue.  Quantum cohomology encodes the OPE ring of the
A model.  Such ideas appear elsewhere in physics, for example there is
a close analogy with results on four-dimensional gauge theories of
Cachazo-Douglas-Seiberg-Witten \cite{cdsw}:
\begin{center}
\begin{tabular}{c|c}
4d ${\cal N}=1$ $SU(N)$ SYM & 2d susy ${\bf C} {\bf P}^{N-1}$ \\ \hline
$S^N = \Lambda^{3N}$ & $x^N = q$ \\
$W = S(1 \: + \: \log( \Lambda^{3N}/S^N) )$ &
$W = \Sigma( 1 \: + \: \log( \Lambda^N/\Sigma^N) )$
\end{tabular}
\end{center}
In the two-dimensional case, the OPE ring looks like a modification of the
classical cohomology ring relation, hence the term ``quantum cohomology.''

In the case of the ${\bf C} {\bf P}^n$ model, the quantum cohomology ring 
corresponds to correlation functions of the form
\begin{displaymath}
< x^k > \: = \: \left\{ \begin{array}{cl}
   q^m &  \mbox{ if }k = mN + N-1 \\
   0 & \mbox{else}.
   \end{array}  \right.
\end{displaymath}
Now, ordinarily for a well-behaved OPE ring, it is assumed that one
needs (2,2) supersymmetry, but it has been found that that condition can
be relaxed to (0,2) supersymmetry.
This was first conjectured by \cite{abs},
and then \cite{sharpe02a} found strong evidence for this
conjecture in our computations of correlation functions.
Later \cite{ade} found a CFT argument explaining
why a well-behaved OPE ring should exist with only (0,2) supersymmetry.

In particular, the paper \cite{abs} studied a (0,2) theory describing
${\bf P}^1 \times {\bf P}^1$ with gauge bundle given by a deformation
of the tangent bundle.  Using a duality argument, they argued
that the quantum cohomology ring of this theory should be given by
\begin{displaymath}
\begin{array}{rcl}
\tilde{X}^2 & = & \exp(it_2) \\
X^2 \: - \: (\epsilon_1 - \epsilon_2) X \tilde{X} & = &
\exp(i t_1)
\end{array}
\end{displaymath}
(where $\epsilon_1 - \epsilon_2$ parametrizes the distance from the
tangent bundle), which is a deformation of the quantum cohomology
ring of ${\bf P}^1 \times {\bf P}^1$.

In \cite{sharpe02a} we directly computed correlation functions
in this theory, using the technology outlined so far, and we found that
\begin{displaymath}
\begin{array}{rcl}
<\tilde{X}^4> & = & < 1 > \exp(2 i t_2) \: = \: 0, \\
< X \tilde{X}^3 > & = & < ( X \tilde{X} ) \tilde{X}^2 >, \\ 
& = & < X \tilde{X} > \exp(i t_2) \: = \: \exp(i t_2), \\
<X^2 \tilde{X}^2 > & = & < X^2> \exp(i t_2) \: = \:
(\epsilon_1 - \epsilon_2) \exp(i t_2), \\
<X^3 \tilde{X}> & = & \exp(i t_1) \: + \: ( \epsilon_1 - \epsilon_2 )^2 
\exp(i t_2), \\
< X^4 > & = & 
2 (\epsilon_1 - \epsilon_2) \exp(i t_1) \: + \:
( \epsilon_1 - \epsilon_2 )^3 \exp(i t_2 ),
\end{array}
\end{displaymath}
verifying the prediction of \cite{abs}.

More recently, \cite{kg} extended the computation of \cite{sharpe02a} to
include all the bundle moduli.  A comparison of their results to the
work of \cite{abs} showed that to obtain agreement between the
Landau-Ginzburg description and the gauged linear sigma model would require
finding a set of potentially complicated field redefinitions
\cite{kg,ilarion1}.  The work of \cite{ilarion2} pointed out that these
correlators may be efficiently computed on the Coulomb branch of the
GLSM.  Besides proving an efficient route to the quantum cohomology 
ring and explicit correlators, these techniques should be helpful in
elucidating the details of the field redefinitions relating the
GLSM to the dual theory of \cite{abs}.
More recently still, in \cite{gs2} A-twists of heterotic Landau-Ginzburg
models in the same universality classes as the heterotic nonlinear
sigma models studied in other work were studied.  These provide alternative
approaches to the same computations, which often make mathematical tricks
manifest.  

So far we have discussed the (0,2) analogue of the A model.
There is also a (0,2) analogue of the B model.  The ordinary (2,2) B model
gets no nonperturbative corrections at all, whereas the (0,2) B model does
get nonperturbative corrections off the (2,2) locus.
These two twists are related in a simple way:
the (0,2) A twist of the theory with bundle ${\cal E}$ on space $X$
is the same, for trivial field-redefinition reasons, as the (0,2) B twist
of the theory with bundle ${\cal E}^{\vee}$ on space $X$.
This is discussed in more detail in \cite{sharpe02c,ade}.

\section{A potential heterotic swampland and new heterotic CFT constructions}

This section is a summary of the paper \cite{DS}, concerning a potential
heterotic swampland and new heterotic worldsheet CFT constructions.

There has recently
been a lot of interest in the landscape program, which for those
readers not already aware is an attempt to extract phenomenological predictions
by doing statistics on the set of string vacua.

One of the technical challenges in the landscape program is that those
string vacua are counted by low-energy effective field theories, and it is
not clear that all of those have consistent UV completions -- not all of them
may come from an underlying quantum gravity \cite{bankstalk,vafaswamp}.

One potential such problem arises in heterotic $E_8 \times E_8$ strings.
The conventional worldsheet construction builds each $E_8$ using a
( ${\bf Z}_2$ orbifold of a ) set of fermions $\lambda_-$.
\begin{displaymath}
L \: = \: g_{\mu \nu} \partial \phi^{\mu} \partial \phi^{\nu} \: + \: i g_{i
\overline{\jmath}} \overline{\psi}_+^{\overline{\jmath}} D_- \psi_+^i
\: + \: h_{a \overline{b}} \overline{\lambda}^{\overline{b}}_- 
D_+ \lambda_-^a \: + \: \cdots.
\end{displaymath}
The fermions realize a Spin(16) current algebra at level 1, and the
${\bf Z}_2$ orbifold makes the current algebra Spin(16)/${\bf Z}_2$.
The group Spin(16)/${\bf Z}_2$ is a subgroup of $E_8$, and we use it to realize
the $E_8$.

In more detail, the adjoint representation ${\bf 248}$
of $E_8$ decomposes into the adjoint
representation ${\bf 120}$ of Spin(16)/${\bf Z}_2$ 
plus a spinor representation ${\bf 128}$:
\begin{displaymath}
{\bf 248} \: = \: {\bf 120} \: + \: {\bf 128}.
\end{displaymath}
The ${\bf 120}$ arises from the left NS sector and the ${\bf 128}$ from the
left R sector of the ${\bf Z}_2$ orbifold.
We take currents transforming in the adjoint and spinor representations of
Spin(16)/${\bf Z}_2$, and form $E_8$ via commutation relations.
More, in fact:  all $E_8$ degrees of freedom, the entire level 1 Kac-Moody
algebra, are realized in this fashion by Spin(16)/${\bf Z}_2$.

This construction has served us well for many years, but, in order to
describe an $E_8$ bundle with connection, that bundle and connection
must be reducible to Spin(16)/${\bf Z}_2$.
After all, all information is buried in the kinetic term
$h_{\alpha \beta} \lambda_-^{\alpha} D_+ \lambda_-^{\beta}$,
and since the $\lambda_-$ transform as a vector representation of
Spin(16), the connection is necessarily only an adjoint of Spin(16).

So, can an $E_8$ bundle with connection always be reduced to
Spin(16)/${\bf Z}_2$ ?
Very briefly:
\begin{itemize}
\item Bundles:  in dimension 9 or less, an $E_8$ bundle can be reduced
to a Spin(16)/${\bf Z}_2$ bundle.
\item Connections:  just because the bundle can be reduced does not mean
the connection on the bundle can be reduced.  We shall find that connections
are not always so reducible -- in fact, this is the generic case.
\end{itemize}
So, is there a heterotic swampland, populated by theories with $E_8$ bundles
with connection which cannot be described in the standard construction?

Let us review the technical issue with connections\footnote{The work on
reducibility of connections was done in collaboration with
R.~Thomas.}; readers interested
in learning about reducibility of bundles are referred to
\cite{DS}.

On a principal $G$ bundle, even a trivial principal $G$ bundle, one can
find connections with holonomy that fill out all of $G$, and so cannot be
understood as connections on a principal $H$ bundle for $H$ a subgroup of $G$:
just take a connection whose curvature generates the Lie algebra of $G$.

Now, in heterotic strings, we cannot work with arbitrary connections,
but rather we want gauge fields satisfying both the anomaly-cancellation
condition $\mbox{ch}_2({\cal E}) = \mbox{ch}_2(TX)$
as well as the Donaldson-Uhlenbeck-Yau condition $g^{i \overline{\jmath}}
F_{i \overline{\jmath}} = 0$.
We shall see that even after imposing these constraints, there are still
examples of $E_8$ bundles with connection that cannot be reduced
to Spin(16)/${\bf Z}_2$.

\begin{figure}[t]   
\centerline{\includegraphics[width=5.5in,height=4.5in]{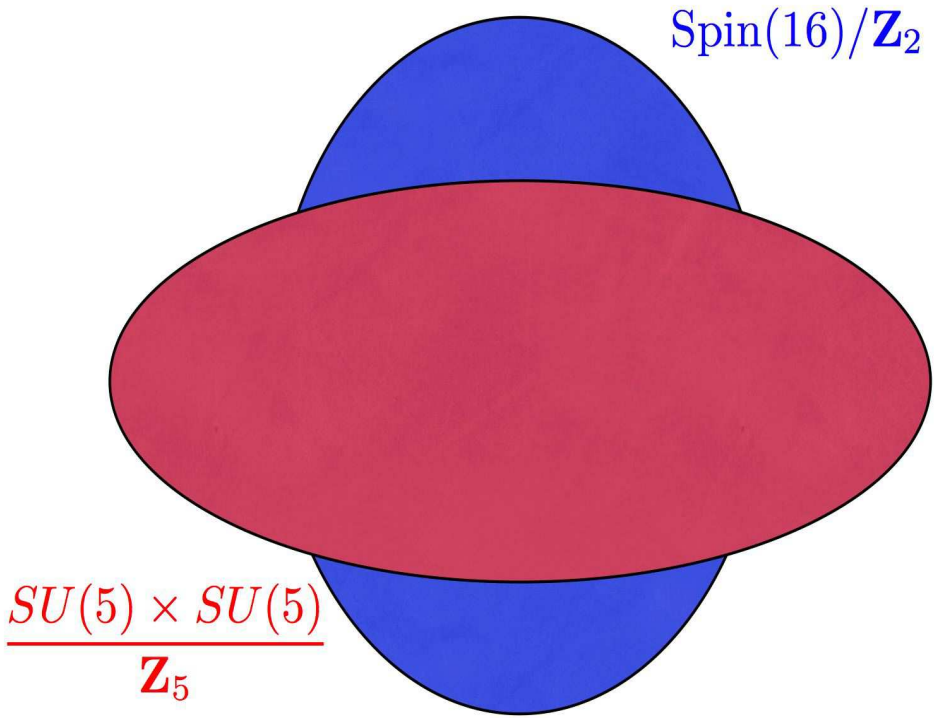}}
\caption{The $\mbox{Spin}(16)/{\bf Z}_2$ and $(SU(5) \times SU(5))/{\bf Z}_5$
subgroups of $E_8$. \label{fig:e8sub}}
\end{figure}

To build such an example, we shall use the fact that $E_8$ has an 
$(SU(5)\times SU(5))/{\bf Z}_5$ subgroup that does not sit inside
Spin(16)/${\bf Z}_2$ (see figure~\ref{fig:e8sub}).  
In particular, $SU(n)$ connections are easy
to build (using holomorphic vector bundles), so we shall construct
an $( SU(5) \times SU(5) )/{\bf Z}_5$ connection that is anomaly-free
and satisfies Donaldson-Uhlenbeck-Yau.

For simplicity, we shall work on an elliptically-fibered $K3$ surface,
and we shall build a stable $SU(5)$ bundle using Friedman-Morgan-Witten
\cite{fmw} technology.
A rank 5 bundle with $c_1=0$, $c_2=12$ has a spectral cover in the
lienar system $|5\sigma + 12 f|$ ($\sigma$ the class of the base, 
$f$ the class of the fiber in the elliptically-fibered surface),
describing a curve of genus $g = 5c_2 - 5^2 + 1 = 36$,
together with a line bundle of degree $-(5+g-1) = -40$.
The result is a (72-parameter) family of stable $SU(5)$ bundles with
$c_2=12$ on $K3$, whose holonomy generically fills out all of $SU(5)$.
If we put two together, and project to the ${\bf Z}_5$ quotient, 
the result is an $(SU(5)\times SU(5))/{\bf Z}_5$ bundle with connection
that satisfies both anomaly cancellation and Donaldson-Uhlenbeck-Yau.

Moreover, this example is not unique -- this describes a 144-dimensional
family of examples.  The lesson here is that such examples are not
exotic special cases, but rather are very common, even generic.

\begin{figure}[t]  
\centerline{\includegraphics[width=5.5in,height=4.5in]{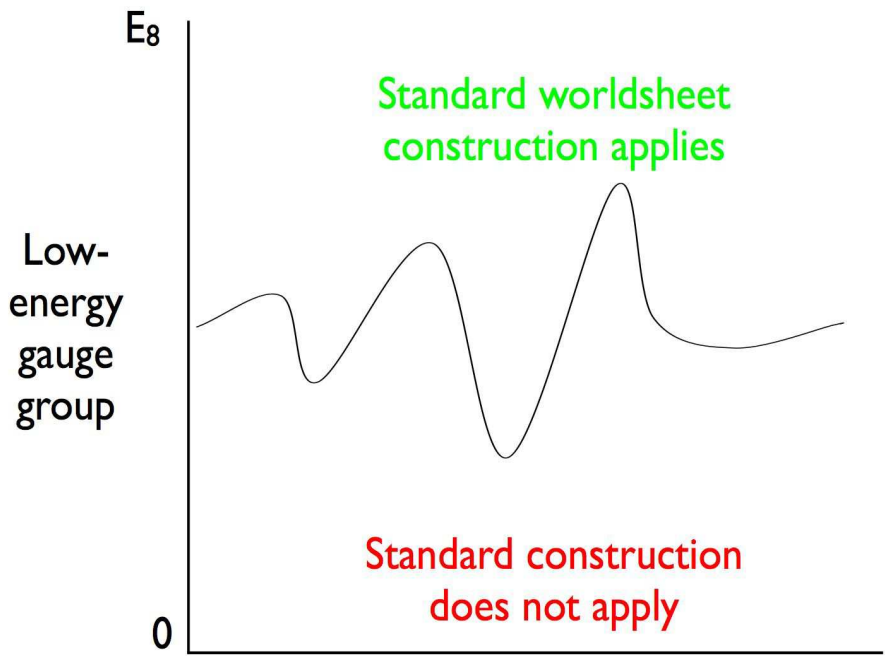}}
\caption{Low-energy gauge groups that can be described by the standard
heterotic worldsheet construction. \label{fig:land}}
\end{figure}

Now that we have set up a problem, we shall see that in fact there is no
swampland, by describing alternative worldsheet constructions that 
can describe more general $E_8$ bundles with connection.
It should be emphasized, however, that even though these more general
$E_8$ gauge fields can be described by some conformal field theory,
there is a danger here for anyone doing statistics on standard
worldsheet constructions -- they do not capture all string vacua,
only a misleading subset, as indicated schematically in figure~\ref{fig:land}.

We shall begin by describing more general ten-dimensional flat-space
constructions.  The basic idea is to replace Spin(16)/${\bf Z}_2$ with
other subgroups of $E_8$, realized as orbifolds of abstract current
algebras (since only $U(n)$ and Spin$(n)$ have level 1 free field
representations).

To be specific, we shall describe $( SU(5) \times SU(5) )/{\bf Z}_5$.

As a first check, let us consider the central charge.
The central charge of a current algebra for an ADE group at level 1
is the same as the rank, so, the central charge of each $SU(5)$ is
4, hence the central charge of an $SU(5) \times SU(5)$ algebra
at level 1 is 8, which will be unchanged by the ${\bf Z}_5$ orbifold.
This is exactly right to match the central charge of $E_8$.

A more convincing test is to study the characters used to build a 
(hopefully modular-invariant) partition function.
In the case of Spin(16)/${\bf Z}_2$, corresponding to the decomposition
\begin{displaymath}
{\bf 248} \: = \: {\bf 120} \: + \: {\bf 128}
\end{displaymath}
of the adjoint representation of $E_8$,
there is a decomposition of characters
\begin{displaymath}
\chi_{E_8}({\bf 1},q) \: = \:
\chi_{Spin(16)}({\bf 1},q) \: + \: 
\chi_{Spin(16)}({\bf 128},q)
\end{displaymath}
which build the left-moving part of the heterotic partition function.
(The adjoint representations are descendants of the identity operator
in level 1 current algebras, and so are buried in the characters for
${\bf 1}$.)

In the case of $SU(5)^2/{\bf Z}_5$, from the decomposition
\begin{displaymath}
{\bf 248} \: = \:
({\bf 1}, {\bf 24}) \: + \: ({\bf 24}, {\bf 1}) \: + \:
( {\bf 5}, {\bf \overline{10} }) \: + \:
( {\bf \overline{5}}, {\bf 10}) \: + \:
( {\bf 10}, {\bf 5}) \: + \:
( {\bf \overline{10}}, {\bf \overline{5}})
\end{displaymath}
of the adjoint representation of $E_8$, we get a prediction for the
characters:
\begin{displaymath}
\chi_{E_8}({\bf 1},q) \: = \:
\chi_{SU(5)}({\bf 1},q)^2 \: + \: 4
\, \chi_{SU(5)}({\bf 5},q) \,
\chi_{SU(5)}({\bf 10},q).
\end{displaymath}
The $SU(5)$ characters can be shown to be
\begin{eqnarray*}
\chi_{SU(5)}({\bf 1},q) & = & \frac{1}{ \eta(\tau)^4} 
\sum_{\vec{m} \in {\bf Z}^4 } q^{ \left( \sum m_i^2 + (\sum m_i)^2 \right)/2}
\\
\chi_{SU(5)}({\bf 5},q) & = & \frac{1}{ \eta(\tau)^4} 
\sum_{\vec{m} \in {\bf Z}^4, \sum m_i \equiv 1 \: mod \: 5 } 
q^{ \left( \sum m_i^2 - \frac{1}{5} (\sum m_i)^2 \right)/2}
\\
\chi_{SU(5)}({\bf 10},q) & = & \frac{1}{ \eta(\tau)^4} 
\sum_{\vec{m} \in {\bf Z}^4, \sum m_i \equiv 2 \: mod \: 5 } 
q^{ \left( \sum m_i^2 - \frac{1}{5} (\sum m_i)^2 \right)/2}.
\end{eqnarray*}

It can be shown \cite{scheidpriv,kacsan} that
\begin{displaymath}
\chi_{E_8}({\bf 1},q) \: = \:
\chi_{SU(5)}({\bf 1},q)^2 \: + \: 4
\, \chi_{SU(5)}({\bf 5},q) \,
\chi_{SU(5)}({\bf 10},q),
\end{displaymath}
which is exactly the desired character identity.

Let us go back and re-examine these statements more carefully.
In the character decomposition for Spin(16)/${\bf Z}_2$, namely
\begin{displaymath}
\chi_{E_8}({\bf 1},q) \: = \:
\chi_{Spin(16)}({\bf 1},q) \: + \: 
\chi_{Spin(16)}({\bf 128},q),
\end{displaymath}
there is a ${\bf Z}_2$ orbifold implicit -- the ${\bf 1}$ is from the
untwisted sector, the ${\bf 128}$ from the twisted sector.
Similarly, in the expression
\begin{displaymath}
\chi_{E_8}({\bf 1},q) \: = \:
\chi_{SU(5)}({\bf 1},q)^2 \: + \: 4
\, \chi_{SU(5)}({\bf 5},q) \,
\chi_{SU(5)}({\bf 10},q),
\end{displaymath}
there is a ${\bf Z}_5$ orbifold implicit -- the ${\bf 1}$ is from
the untwisted sector, the other four pieces are from four twisted sectors.
This is exactly what we should find -- the correct subgroup of $E_8$
is $SU(5)^2/{\bf Z}_5$, not $SU(5)^2$; there is a ${\bf Z}_5$ that
should replace the left-moving ${\bf Z}_2$ GSO-analogue of the ordinary
heterotic string construction.

It is straightforward to repeat this analysis for other subgroups of
$E_8$.  For example, let us next consider $SU(9)/{\bf Z}_3$.
Again, the central charge of the current algebra at level 1 is 8,
which matches that of the $E_8$ current algebra.
From the decomposition of the $E_8$ conformal family
\begin{displaymath}
[{\bf 1}] \: = \: [{\bf 1}] \: + \: [ {\bf 84} ] \: + \:
[ {\bf \overline{84}} ]
\end{displaymath}
one predicts a character identity
\begin{displaymath}
\chi_{E_8}({\bf 1},q) \: = \:
\chi_{SU(9)}({\bf 1},q) \: + \: 
2 \, \chi_{SU(9)}({\bf 84},q)
\end{displaymath}
and it can be shown that this is correct.
Furthermore, note that in the expression above, there is a ${\bf Z}_3$
orbifold implicit, exactly as desired since the correct subgroup is
$SU(9)/{\bf Z}_3$.
So, we can also describe the $E_8$ degrees of freedom with 
$SU(9)/{\bf Z}_3$, with a ${\bf Z}_3$ orbifold replacing the
left-moving ${\bf Z}_2$ GSO-analogue of the ordinary
heterotic string construction.

For many other maximal-rank subgroups, there is an analogous story.
For at least one non-maximal-rank, non-simply-laced subgroup, however,
matters are more complicated.  $E_8$ has a $G_2 \times F_4$ subgroup,
and one can check that the central charge of the $G_2 \times F_4$
current algebra at level 1 is 8, exactly right to match that of $E_8$,
and furthermore there is a character identity relating
$\chi_{E_8}({\bf 1},q)$ to a sum over $G_2 \times F_4$ characters,
as needed.  However, that character sum includes more terms that
just characters of identity representations, terms which ordinarily
would arise in twisted sectors of orbifolds, but there is no
group-theoretic orbifold to perform:  $G_2\times F_4$ has
no finite center or indeed a finite normal subgroup that
one could orbifold, and correspondingly the subgroup of $E_8$ is
$G_2 \times F_4$, not any quotient thereof.
Therefore, although there exists formally a character decomposition,
we do not understand how to realize it physically.

Next, in order to make this useful, we must describe how one fibers
more general current algebras over spaces, in order to describe
a compactified theory.  To do this, we will outline
fibered WZW models, first introduced by 
\cite{gates1,gates2,gates3,gates4,gates5}
under the name `lefton, righton Thirring models,'' which we specialize
to (0,2) theories.

Before describing fibered WZW models, let us first recall ordinary WZW models.
An ordinary WZW model looks like a sigma model on the group manifold of 
a group $G$ together with $H$ flux:
\begin{displaymath}
S \: = \: - \frac{k}{2 \pi} \int_{\Sigma}
\mbox{Tr }\left[ g^{-1} \partial g g^{-1}
\overline{\partial} g \right]
\: - \: \frac{i k }{2 \pi} \int_B d^3 y \epsilon^{ijk}\mbox{Tr }\left[ g^{-1}
\partial_i g g^{-1} \partial_j g
g^{-1} \partial_k g \right].
\end{displaymath}
The first term is the usual sigma model, the second or `Wess-Zumino' term
describes $H$ flux.
This theory has a $G_L \times G_R$ symmetry, with currents
\begin{displaymath}
J(z) \: = \: g^{-1} \partial g, \: \: \:
\overline{J}(\overline{z}) \: = \: 
\overline{\partial} g \, g^{-1}
\end{displaymath}
which (thanks to the Wess-Zumino term) obey
\begin{displaymath}
\overline{\partial} J(z) \: = \:
\partial \overline{J}(\overline{z})
\: = \: 0.
\end{displaymath}
These currents each realize a chiral $G$-Kac-Moody or current algebra
at level $k$.

Next, let us describe how to fiber these WZW models.
Let $P$ be a principal $G$ bundle over $X$, with connection $A$.
Given an ordinary heterotic string worldsheet theory, the idea is that
we shall replace the left-moving fermions $\lambda_-$ with 
a WZW model with left-multiplication gauged with the pullback of $A$.
Schematically, the action has the form
\begin{eqnarray*}
\lefteqn{
\frac{1}{\alpha'} \int_{\Sigma} \left(
g_{i \overline{\jmath}} \partial_{\alpha}
\phi^i \partial^{\alpha} \phi^{\overline{\jmath}} \: + \: \cdots \right)
} \\ \\
& & \: \: - \: \frac{k}{4 \pi} \int_{\Sigma} 
\mbox{Tr }\left( g^{-1} \partial g g^{-1} \overline{\partial} g \right) 
\: - \: \frac{i k}{12 \pi} \int_B d^3 y \epsilon^{ijk} \mbox{Tr }
\left( g^{-1} \partial_i g g^{-1} \partial_j g g^{-1} \partial_k g \right) \\
& & \: \: - \: \frac{k}{2 \pi} \mbox{Tr }\left( ( \partial \phi^{\mu}) 
A_{\mu} \overline{\partial} g g^{-1} \: + \: \frac{1}{2} (
\partial \phi^{\mu} \overline{\partial} \phi^{\nu} ) A_{\mu} A_{\nu} \right).
\end{eqnarray*}
The first line describes a nonlinear sigma model on $X$;
the second line describes a WZW model;
the third line describes the operation of gauging left-multiplication.
The third line will look very familiar to readers acquainted with gauged WZW
models; the differences here are that we only gauge a chiral 
multiplication, and that the gauge field is the pullback of a gauge
field on the target space, rather than a worldsheet gauge field.

Now, a WZW model action is invariant under gauging symmetric group 
multiplications,
but not under the chiral group multiplications that we have here.
Under
\begin{eqnarray*}
g & \mapsto & h g \\
A_{\mu} & \mapsto & h A_{\mu} h^{-1} \: + \: h \partial_{\mu} h^{-1}
\end{eqnarray*}
the classical action is not invariant.

On the one hand, this lack of invariance is expected -- this is the bosonization
of the chiral anomaly.
On the other hand, this lack of invariance creates a potential well-definedness
issue in the fibered WZW construction.

The fix is a quantum correction which cancels the classical non-invariance.
In particular, the classical action picks up a quantum correction across
coordinate
patches, due to the right-moving chiral fermion anomaly.

To make the action gauge-invariant, we proceed in the usual form,
by assigning a transformation law to the $B$ field.
To be able to do this globally implies that
\begin{displaymath}
k \, \mbox{ch}_2({\cal E}) \: = \: \mbox{ch}_2(TX)
\end{displaymath}
which is the analogue of the usual anomaly-cancellation condition for
left-movers at level $k$.
If this condition is obeyed, the action is well-defined globally.

Proceeding in the usual fashion for heterotic strings, the right-moving
fermion kinetic terms on the worldsheet couple to $H$ flux:
\begin{displaymath}
\frac{i}{2} g_{\mu \nu} \psi_+^{\mu} D_{\overline{z}} \psi_+^{\nu}
\end{displaymath}
where
\begin{displaymath}
D_{\overline{z}} \psi_+^{\mu} \: = \: \overline{\partial} \psi_+^{\mu} \: + \: \overline{\partial} \phi^{\mu} \left( \Gamma^{\nu}_{\: \: \sigma \mu} \: - \:
H^{\nu}_{\: \: \sigma \mu} \right) \psi_+^{\sigma}.
\end{displaymath}
To make the fermion kinetic terms gauge-invariant under the
transformation of the $B$ field, we must redefine $H$
to be
\begin{displaymath}
H \: = \: dB \: + \: (\alpha')\left(
k CS(A) \: - \: CS(\omega) \right)
\end{displaymath}
so that $H$ is a well-defined 3-form globally.
This implies that
\begin{displaymath}
k \, \mbox{ch}_2({\cal E}) \: = \: \mbox{ch}_2(TX)
\end{displaymath}
which is, again, the anomaly cancellation condition.

Next, let us demand (0,2) supersymmetry.
One discovers an old faux-supersymmetry-anomaly in subleading terms in
$\alpha'$ in the heterotic string, one originally discussed by
\cite{sen1,sen2}.
The supersymmetry transformations in the ordinary heterotic string
worldsheet are:
\begin{displaymath}
\delta \lambda_- \: = \: - i \epsilon \psi_+^{\mu} A_{\mu} \lambda_-.
\end{displaymath}
Note that this is the same as a chiral gauge transformation with
parameter $ - i \epsilon \psi_+^{\mu} A_{\mu} $.
Because of the chiral anomaly in the ordinary heterotic worldsheet,
this means there is a quantum contribution to the supersymmetry transformations
at order $\alpha'$.  In our bosonized description, this quantum effect
appears at leading order.

The one-fermi terms in the supersymmetry transformations of the various
components of the action are as follows.
The nonlinear sigma model on the base contributes
\begin{displaymath}
\frac{1}{\alpha'} \int_{\Sigma} (i \alpha \psi^{\overline{\imath}} ) \overline{\partial} \phi^{\mu} \partial \phi^{\nu}
\left( H
\: - \: d B \right)_{\overline{\imath} \mu \nu}.
\end{displaymath}
The WZW fibers contribute
\begin{displaymath}
- k\int_{\Sigma} (i \alpha \psi^{\overline{\imath}}) \overline{\partial} \phi^{\mu} \partial \phi^{\nu}
CS(A)_{\overline{\imath} \mu \nu}.
\end{displaymath}
This is the bosonization of the faux-supersymmetry-anomaly discussed above.
Finally, there is an additional quantum contribution, a 
faux-supersymmetry-anomaly arising from the right-moving fermions,
which contributes
\begin{displaymath}
\int_{\Sigma} (i \alpha \psi^{\overline{\imath}}) \overline{\partial} \phi^{\mu} \partial \phi^{\nu}
CS(\omega)_{\overline{\imath} \mu \nu}.
\end{displaymath}
In order for the one-fermi terms to cancel out, we see that
\begin{displaymath}
H \: = \: dB \: + \: \alpha'\left( k CS(A) \: - \: CS(\omega) \right)
\end{displaymath}
which is, again, the anomaly cancellation condition,
the third time now that we have derived it.
The three-fermi terms behave similarly.

Next, let us turn to the massless spectrum, or more properly the spectrum
of chiral primaries.
In an ordinary WZW model, the WZW primaries are associated to integrable
representations of the group $G$.
Here, for each integrable representation $R$ of the principal $G$ bundle $P$,
we get an associated vector bundle ${\cal E}_R$.
The chiral primaries are then easily checked to be given by
$H^*(X, {\cal E}_R)$ for each integrable representation $R$.

For an example, consider $G=SU(n)$ at level 1.
Here, the integrable representations are the fundamental ${\bf n}$ and
its exterior powers, hence the chiral primaries are counted by
$H^*(X, \Lambda^* {\cal E})$, in agreement with old results
\cite{dg}.

These fibered WZW constructions realize the `new' elliptic genera
of Ando and Liu \cite{kliu,ando1}.
Ordinary elliptic genera describe left-movers coupled to a level 1
current algebra; these elliptic genera, on the other hand, have 
left-moving level $k$ current algebras.

It would be very interesting if these elliptic genera could be applied
to understand, for example, black hole entropy computations.

\section{String compactifications on stacks}

In this section we will review results in
\cite{PS0,PS1,PS2,PS3,ESDCStx,glsmhpd}, describing string compactifications
on stacks and applications to gauged linear sigma models.

Stacks are a mild generalization of spaces,
on which there exist functions, metrics, spinors, and all the other
machinery one needs to make sense of strings.

One would like to understand string compactifications on stacks,
not only to understand the most general possible string compactifications,
but also because they often appear physically inside various constructions.

So, what is a stack?
We can cover a stack with coordinate charts, just like a manifold.  
So, locally, a stack looks just like a space.
Globally, on a manifold, across triple overlaps, the coordinate
charts close to the identity.  On a stack, across triple overlaps, the
coordinate charts need only close up to an automorphism.
In other words, locally a stack is just like a space -- the difference
is global.

How does one make sense of string compactifications on stacks concretely?
Every\footnote{With minor caveats.} smooth, Deligne-Mumford stack can be
presented as a global quotient $[X/G]$ for $X$ a space and $G$ a group.
To such a presentation, we associate a $G$-gauged sigma model on $X$.

Following that program, we quickly run into a potential problem.
If to $[X/G]$ we associate a $G$-gauged sigma model, then
\begin{itemize}
\item $[{\bf C}^2/{\bf Z}_2]$ defines a two-dimensional conformal field
theory,
\item $[X/{\bf C}^{\times}]$ for $X = \left( {\bf C}^2 \times {\bf C}^{\times}
\right)/{\bf Z}_2$ is an isomorphic stack, but $[X/{\bf C}^{\times}]$
defines a two-dimensional theory without conformal invariance.
\end{itemize}
Thus, there is a potential presentation-dependence problem:
the same stack, presented in two different ways, is associated with
distinct quantum field theories.  Just like in the physical realization
of derived categories, wehre an analogous problem arises,
our proposal is that this presentation-dependence is fixed by
renormalization-group flow.
In other words, stacks classify endpoints of renormalization-group flow.

There are some potential problems with this program, reasons to believe
that presentation-independence fails.
One example involves deformation theory.  The first check of a geometrical
interpretation of any physical structure is to compare physical
deformations to mathematical deformations, and in all previously-known cases,
they match.  For stacks, on the other hand, there is a mismatch:
deformations of physical theories do not match mathematical deformations
of stacks.
Another example of a potential problem with the program outlined
above involves cluster decomposition, or rather lack thereof,
in the theories one associates to special kinds of stacks known as gerbes.

These potential problems can be fixed \cite{PS0,PS1,PS2,PS3,ESDCStx}.
The results include:  mirror symmetry for stacks, new Landau-Ginzburg
models, physical calculations of quantum cohomology for stacks,
and an understanding of noneffective quotients in physics.

For example, let us consider quantum cohomology.
As outlined earlier, the quantum cohomology ring of ${\bf C} {\bf P}^N$
is ${\bf C}[x]/(x^{N+1} - q)$, a deformation of the classical
cohomology ring ${\bf C}[x]/(x^{N+1})$.
The quantum cohomology ring of a ${\bf Z}_k$ gerbe over
${\bf C}{\bf P}^N$ with characteristic class $-n \mbox{ mod }k$
is
\begin{displaymath}
{\bf C}[x,y]/\left( y^k - q_2, x^{N+1} - y^n q_1 \right).
\end{displaymath}
In particular, even though the gerbe is not a space, one can still make
sense of notions familiar from ordinary spaces.

For another example, the Toda dual of ${\bf C} {\bf P}^N$ is described
by the holomorphic function
\begin{displaymath}
W \: = \: \exp(-Y_1) \: + \cdots \:
+ \: \exp(-Y_N) \: + \: \exp(Y_1 \: +
\: \cdots \: + Y_N).
\end{displaymath}
The analogous duals to ${\bf Z}_k$ gerbes over ${\bf C} {\bf P}^N$ are
described by
\begin{displaymath}
W \: = \: \exp(-Y_1) \: + \cdots \:
+ \: \exp(-Y_N) \: + \: \Upsilon^n \exp(Y_1 \: +
\: \cdots \: + Y_N)
\end{displaymath}
where $\Upsilon$ is a character-valued field \cite{PS0,PS1,PS2}.

More generally, there exists a notion of toric stacks \cite{bcs}
which allows us to realize sigma models on many stacks in terms of simple
two-dimensional gauge theories.
Standard mirror constructions now produce discrete-valued fields
\cite{PS0,PS1,PS2}, a new
effect, which ties into the stacky fan description of \cite{bcs}.

We now believe that for banded, abelian $G$-gerbes on spaces $X$,
(2,2) supersymmetric strings cannot distinguish between
\begin{itemize}
\item the gerbe itself,
\item disjoint union of copies of $X$ (one for each character of $G$), 
with flat $B$ fields,
determined by the image of the map
\begin{displaymath}
H^2(X,G) \: \stackrel{ G \rightarrow U(1) }{ \longrightarrow } \:
H^2(X, U(1)).
\end{displaymath}
\end{itemize}
In other words, these are described by the same conformal field theory.
This fact comes up in mirror constructions, and the meaning of discrete-valued
fields, and has impliciations for quantum cohomology computations
\cite{PS3}.

This result can also be applied to understand gauged linear sigma models.
For example, let us consider the GLSM for the complete intersection
${\bf P}^7[2,2,2,2]$.
At the Landau-Ginzburg point, onehas a superpotential
of the form
\begin{displaymath}
\sum_a p_a G_a(\phi) \: = \:
\sum_{ij} \phi_i A^{ij}(p) \phi_j
\end{displaymath}
where the $G_a$ are quadric polynomials, and the $A^{ij}(p)$ is
an $8 \times 8$ matrix with entries linear in the $p$'s,
and the $\phi_i$ are the homogeneous coordinates on ${\bf P}^7$.
In particular, this superpotential describes mass terms for
the $\phi_i$, away from the locus $\{ \det A = 0 \}$,
leaving just the $p$ fields.
Since the $p$ fields have charge $-2$, this describes a ${\bf Z}_2$
gerbe, which physics sees as a double cover.
Altogether, the Landau-Ginzburg point describes a branched double cover
of ${\bf P}^3$.
This is a non-birational twisted derived equivalence, and a physical
realization of Kuznetsov's homological projective duality
\cite{glsmhpd}.

Although (2,2) models on gerbes decompose into a disjoint union,
(0,2) models do not in general.  Heterotic strings on gerbes with
twisted bundles do not factorize into a disjoint union of target spaces.
The prototype for such bundles is the ``${\cal O}(1/k)$'' line bundle
over the ${\bf Z}_k$ gerbe ${\bf P}^N_{[k, k, \cdots, k]}$
on ${\bf P}^N$, defined by a Fermi superfield of (minimal) charge 1.
More generally, heterotic strings on gerbes give an understanding of
some of the two-dimensional (0,4) theories appearing in the geometric
Langlands program.  Implicit here is a lesson for the landscape program:
many more string vacua may exist than previously enumerated.

\section{Conclusions}

In this talk we have outlined three recent developments pertinent to
heterotic strings.  First, we discussed recent progress in understanding
nonperturbative corrections in heterotic strings.  Second, we discussed
a potential heterotic swampland -- the inability of ordinary heterotic
worldsheet constructions to describe most $E_8$ bundles with connection --
and its resolution via the construction of new heterotic worldsheet CFT's.
Finally, we outlined string compactifications on stacks.
These compactifications give new insight into many gauged linear sigma models,
ranging from exotic-seeming GLSM's with nonminimal charges, to
seemingly ordinary GLSM's which have novel geometries appearing at various
limits of their K\"ahler moduli spaces, physically realizing Kuznetsov's
homological projective duality and Kontsevich's noncommutative spaces.  
Heterotic string compactifications
on special stacks known as gerbes appear to provide a new class of
string compactifications.

\end{document}